\documentclass[twocolumn,showpacs,preprintnumbers,amsmath,amssymb,superscriptaddress,jap]{revtex4-1}
\usepackage{graphicx}
\usepackage{dcolumn}
\usepackage{bm}
\usepackage{comment}

\begin{document}


\title{ Quantum logic gates  from  time-dependent global  magnetic field in a system  with constant exchange  }

\author{A.~V.~Nenashev}
\email {nenashev@isp.nsc.ru}\affiliation{Rzhanov Institute of
Semiconductor Physics SB RAS, 630090 Novosibirsk, Russia}
\affiliation{Novosibirsk State University, 630090 Novosibirsk,
Russia}

\author{A.~F.~Zinovieva}
\affiliation{Rzhanov Institute of Semiconductor Physics SB RAS,
630090 Novosibirsk, Russia}

\author{A.~V.~Dvurechenskii}
\affiliation{Rzhanov Institute of Semiconductor Physics SB RAS,
630090 Novosibirsk, Russia} \affiliation{Novosibirsk State
University, 630090 Novosibirsk, Russia}

\author{A.~Yu.~Gornov}
\author{T.~S.~Zarodnyuk}
\affiliation{Institute for System Dynamics and Control Theory SB
RAS, 664033 Irkutsk, Russia}

\date{\today}

\begin{abstract}
We propose a method for implementation of an universal set of one-
and two-quantum-bit gates for quantum computation in the system of
two coupled electrons with constant non-diagonal exchange
interaction. Suppression of the exchange interaction is offered to
implement by all-the-time repetition of single spin rotations. Small
g-factor difference of electrons allows to address qubits and to
avoid strong magnetic field pulses. It is shown by means of
numerical experiments that for implementation of one- and two-qubit
operations it is sufficient to change the amplitude of the magnetic
field within a few Gauss, introducing in a resonance one and then
the other electron. To find the evolution of the two-qubit system,
we use the algorithms of the optimal control theory.
\end{abstract}


\maketitle

\section*{Introduction}
There have been numerous proposals for implementation of quantum
computation schemes in different material realizations, such as
photon qubits,  trapped atoms and ions, nuclear spins in molecules
in liquid solutions, spin or charge states in quantum dots (QDs) or
dopants in solids, and superconducting circuits \cite{a1}. Each such
system must satisfy to 5 DiVinchenzo criteria: scalability,
initialization ability, long coherence lifetime,  universal  set of
quantum gates realization,  readout ability \cite{a2}.  Inspired by
the Loss and DiVincenzo proposal \cite{a3}, there is a continuing
experimental effort to realize electron spin quantum bits or qubits
in semiconductor quantum dots.\cite{a4} There, a single qubit is
defined by the two spin states of an electron localized in a quantum
dot (QD).  Coupling between qubits is provided by exchange
interaction between electrons in neighboring quantum dots. For
single-qubit rotations the electron spin resonance (ESR) technique
was proposed \cite{a5}. Two-qubit operations can be performed as
combinations of the $\sqrt{SWAP}$ operation (based on the exchange
interaction) with single-spin rotations. Read-out procedure can be
realized via the spin-to-charge conversion \cite{a6}. In light of
the foregoing, semiconductor QD systems are well suited for quantum
computing, provided that the decoherence time in the system is long
enough and there is a possibility of the system initialization.
Realization of this proposal needs a system of electrodes. One type
of electrodes provides the control of the  exchange coupling during
two-qubit operations, and the second type of electrodes allows  to
address an individual spin for one-qubit operations. In subsequent
articles, the g-factor engineering or the local magnetic field
gradient were proposed additionally for one-qubit operations
\cite{a7,a8}.

Two-qubit manipulation using electrical gates is actually quite a
challenge. The complexity of this approach can be recognized by
comparison of two following dates, the Loss and DiVincenzo proposal
\cite{a3} has been published in 1998, while an experimental
implementation of two-qubit operation in a double quantum dot has
been made only in 2011 (see Ref.~\onlinecite{a8}). All used
electrodes can be sources of fluctuations and can lead to undesired
errors. The solution of this problem is to exclude the gates at
least partially, for example, to refuse electrodes controlling the
exchange interaction and to find a way of the coupling control
without electrodes.

There are some works where authors proposed quantum computation
schemes based on the constant exchange coupling. The exchange
coupling was eliminated by using encoded bits \cite{a9,a10}, where
two physical qubits in the states $\uparrow$ and $\downarrow$ are
unified in one logical qubit $\uparrow\downarrow$. This
configuration provides  compensation of interactions with the
environment. But two-qubit operations demand leaving the
interaction-free space. By a strong local magnetic field the state
$\uparrow\downarrow$ transforms to $\uparrow\uparrow$, and the
interaction between logical qubits becomes switched on \cite{a9}.
After performing the two-qubit operation the qubits are driven back
to the interaction-free space. One more approach for elimination of
the exchange interaction is based on refocusing pulses that are used
in nuclear magnetic resonance (NMR) technique [Sec.~7.7.3 in
Ref.~\onlinecite{a11}]. It is applicable if the inter-qubit
interaction Hamiltonian is diagonal in the computational basis. The
repetition of refocusing pulses allows to suppress the interaction
between qubits. Refocusing requires very fast repeated switchings
with a period much shorter than the elementary operation time, that
is quite difficult to implement in practice. Another complexity is
locality of refocusing pulses, it requires injecting pulses onto
every qubit of the quantum system. The use of strong local magnetic
fields or the refocusing technique requires a more complicated
architecture of the quantum computer, and we have a goal to find
another way to eliminate the coupling between qubits.

An idea proposed in the present work is close to the one of the work
of Ozhigov and Fedichkin, \cite{a12} where elimination of the
exchange interaction by all-the-time repetition of one-qubit
operations was proposed. We offer repeating single-spin rotations
for suppression of the exchange interaction, when the latter must be
switched off. Our approach allows not only elimination of the
exchange interaction, but also implementation of all basic logical
operations.

This paper is organized as follows. In Sec.~I we describe the
intuitive scheme of our proposal, that is represented as a
mathematical model in Sec.~II. In Sec.~III the ways of
implementation of basic logical operations are discussed in the
frame of this model. In Sec. IV the possibility of one- and
two-qubit operations implementation is demonstrated by means of
numerical experiments.  The evolution of qubit states  is found
using the algorithms of the optimal control theory \cite{a13}.
Optimization was performed using a combination of nonlocal search
methods and a local descent by conjugate gradient methods.

\section{Main idea}

The intuitive scheme of our proposal is the following. We take two
electrons with a constant exchange coupling between them and provide
all-the-time rotation of their spins by application of the resonance
microwave frequency radiation. These spin rotations eliminate the
exchange interaction between electrons. For two-qubit gates
implementation we switch on the exchange coupling taking electrons
out of resonance. This can be done by simple addition of a small
magnetic field $B\pm\delta B(t)$. The most important two-qubit
operation CNOT can be done through the combination of single spin
rotations and the $\sqrt{SWAP}$ operation. For the single-spin
rotations electrons are driven to the resonance, and for the
$\sqrt{SWAP}$ operation electrons are taken out of the resonance.
Single spin rotations (when one electron spin of the coupled pair
rotates, and the second one does not rotate) need the difference in
electron g-factors. If this difference would be not so large (for
example, $\delta g\sim10^{-3}$) one can perform quantum operations
by a small addition to the magnetic field $\delta B(t)\sim1$~G, that
can be produced by conventional coils used in NMR technique
\cite{a14}. Such a difference $\delta g$ can be realized by g-factor
engineering in various quantum dot or quantum well systems,  by
changing the alloy composition in A$_3$B$_5$ \cite{a15,a16}, or due
to localization of electrons in different energetic valleys in a
Ge/Si QD system \cite{a17}. A Ge/Si system with quantum dots seems
to be more suitable, because such a pair of electrons can be
localized at one quantum dot, one at the apex of the quantum dot,
and a second electron at the QD base edge \cite{a17}.

The essential difference of our approach from previous schemes of
quantum computation is the use of the g-factor difference of two
electrons for effective control (switching on/off) of the exchange
interaction between them. In existing works the g-factor difference
is used for addressing individual qubits and implementation of
single-qubit operations. There have been proposed very original
schemes of single-qubit gate implementation, using the g-factor
difference, but without conventional microwave pulses
\cite{a18,a19}. For example, in the work of Levy \cite{a18} periodic
modulation of the exchange coupling within logical qubit at the Rabi
frequency $\Omega=\Delta g B$ is proposed to produce $\pi$ and
$\pi/2$ pulses.  In some works the g-factor difference is used in
spin-charge conversion schemes. For example, in the work of Yokoshi
{\it et al.} \cite{a20}  sequential spin-to-charge conversions allow
to implement the full Bell state measurement of electron-spin
qubits. However, in all these proposals the exchange coupling must
be controlled by electrodes. Our approach allows the implementation
of single-qubit and two-qubit gates without electrodes controlling
the exchange within one operational unit (an electron pair localized
at one QD).

For large-scale quantum computation at least  1000 qubits are
needed. It is possible to create a large ordered array of quantum
dots by epitaxial growth on pre-patterned substrates \cite{a21}. Let
each quantum dot possess a  pair of  electrons with a constant
coupling and different g-factors \cite{a17}, then it can be
considered as a basic element for logical gates implementation.
Between quantum dots one can organize the $SWAP$ operations  by
means of additional electrodes created on the surface. Since all
quantum dots are grown in one and the same growth manner, they are
almost identical. To provide the selective access to an individual
QD we suggest creating a parallel layer of smaller QDs (a storage
layer), where two small quantum dots correspond to one large QD in
the first layer (Fig.~1). Let electrons localized on smaller QDs
have the g-factor quite different from g-factors of electrons in the
first layer, then they are always out of resonance, and these
quantum dots can serve as storage elements for electron spins. In
the idle mode electrons are located in the storage layer. The
similar structures were created using the strain-induced nucleation
of QD molecules above buried nanomounds \cite{a22}. For performing
logic gates the electrons from a pair of QDs in the storage layer
are pushed to the corresponding QD in the upper (operation) layer by
the ``Address'' electrode placed above the selected QD (see Fig.~1).
The similar idea is described in Ref.~\onlinecite{a7} where
electrons are driven by electrodes away from the dopant ion into
layers of different alloy composition (and respectively different
g-factors) for implementation of one- and two-qubit operations.

\begin{figure}
\includegraphics[width=3.2in]{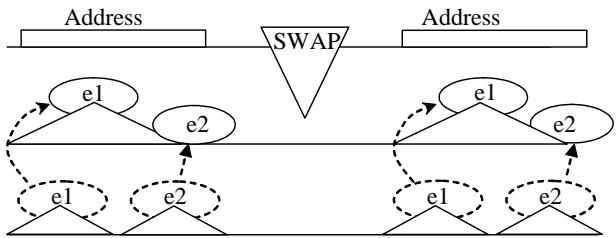}
\caption{\label{f2} Arrangement of quantum dots in the active
(upper) layer and the storage (lower) layer, designed to perform
quantum logic operations based on the control of the exchange
interaction  by means of the single spin rotations.}
\end{figure}

Our approach has the following advantages: it allows (i)  to remove
a half of electrodes controlling the exchange interaction, namely
the electrodes controlling the exchange within electron pairs, (ii)
to refuse the microwave pulse technics, (iii) to refuse the local
strong magnetic field, because both magnetic fields $B$ and $\delta
B(t)$ used in our approach are global, (iv) to use standard
 ESR and NMR techniques.

In this work we verify the possibility of implementation of basic
logical operations on the example of the Ge/Si system with quantum
dots. Ge QDs grown at certain growth conditions   allow to localize
two electrons on the same QD: near the QD apex and near the QD base
edge (Fig.~1), that provides different g-factors due to localization
in different $\Delta$ valleys, namely $g_{||}=1.9995$ and
$g_{\perp}=1.9984$ \cite{a17}. Being close to each other, these
electrons can have a sufficient overlap of wave functions required
for two-qubit operations. One more advantage of the Ge/Si
heterosystem is the location of electrons in Si, that can lead to
long spin relaxation times due to small spin-orbit interaction in
this material and small concentration of the $^{29}$Si isotope with
a nonzero nuclear spin \cite{a23,a24}. Direct measurements of
electron spin lifetimes in dense arrays of Ge/Si QDs give the times
of the order of 10~$\mu$s \cite{a25}. The spin lifetimes can be
increased by switching off the main mechanism of spin relaxation in
Ge/Si QD system---stochastic spin precession at the tunneling
between QDs \cite{a26}. Indeed, partial suppression of the tunneling
provides fourfold increase of spin lifetimes \cite{a27}. Further
increase of spin lifetimes  can be reached by using the isotopically
pure material $^{28}$Si \cite{a28}. In general, other heterosystems,
such as $A_3B_5$,  can be used for realization of our proposal.

\section{Model}
Let us represent the proposed model in a mathematical form. We
consider two electrons with a constant tunneling coupling  in the
magnetic field (see Fig.~2). Let these electrons have close
$g$-factors, differing on the small magnitude, $g_1=g_0-\delta g/2$
and $g_2=g_0+\delta g/2$ with $\delta g\ll g_0$.  A constant
magnetic field $B_0$ is applied in the $z$-direction, providing
Larmor precession of the electron spins. This precession is
described by the term
\[
\hat H_{0}=\mu_B B_0 (g_1 \hat S_{1z} + g_2 \hat S_{2z}),
\]
where $\mu_B$ is the Bohr magneton, $g_1$ and $g_2$ are $g$-factors
of the electrons, $\hat{\mathbf{S}}_1$ and $\hat{\mathbf{S}}_2$ are
their spin operators.

Electrons are subjected to a microwave radiation circularly
polarized in the $xy$-plane. The magnetic field $\mathbf{B}_w(t)$ of
the microwave is
\[
\mathbf{B}_m(t) = (B_{m_0} \cos \Omega t, B_{m_0} \sin \Omega t, 0),
\]
where $B_{m}$ is the amplitude of the microwave field, and $\Omega$
is the frequency. The latter is chosen to be equal to the mean value
of Larmor frequencies of two electrons:
\[
\hbar \Omega = \frac{g_1+g_2}{2} \mu_B B_0 ,
\]
thus both electrons in the stationary conditions are out of the spin
resonance. The corresponding term in the Hamiltonian is
\[
\hat H_{m}(t) = \mu_B \mathbf{B}_m(t) (g_1 \hat{\mathbf{S}}_1 + g_2
\hat{\mathbf{S}}_2).
\]

These electron spins are considered as qubits. The model includes
the interaction between qubits having a form of the Heisenberg
exchange interaction:
\[
\hat H_{int}=J\hat{\mathbf{S}}_1\hat{\mathbf{S}}_2,
\]
where $J$ is the exchange integral, which is supposed to be
constant.

The main idea  is to make the system controllable with an addition
of a small time-dependent magnetic field $\delta B(t)$. This
magnetic field is also directed along the axis $z$, and the
corresponding term in the Hamiltonian is
\[
\hat H_{c}(t) = \mu_B \delta B(t) (g_1 \hat S_{1z} + g_2 \hat S_{2z}).
\]
By means of this small additive it is possible to enter one of
electrons into the resonance  for a required time,  and  to carry
out the spin turning on a desired angle through Rabi oscillations.
Changing the additive $\delta B(t)$ one can rotate the spin of
another electron. In this way it is possible to manipulate with
qubits in this system and to carry out the main quantum logical
operations.

The spin behavior becomes simpler in the  reference frame, rotating
with the frequency $\Omega$.  In this frame, the main contribution
to the spin precession is removed and only relatively slow dynamics
remains. The full Hamiltonian of the system of two coupled
electrons, written in the rotating frame, takes the following form:

\[
\hat H(t) = \hat H_0 +\hat H_m(0) +\hat H_{int} +\hat H_c(t)
-\hbar\Omega( \hat S_{1z} + \hat S_{2z}).
\]

In the basis $|\!\!\uparrow\uparrow\rangle$,
$|\!\!\uparrow\downarrow\rangle$, $|\!\!\downarrow\uparrow\rangle$,
$|\!\!\downarrow\downarrow\rangle$ this Hamiltonian has the
following matrix representation:
\begin{widetext}
$$\hat H(t) = \left ( \begin {array}{cccc}
\frac{J}{4}+g_0\mu_B\delta B(t) & \frac{g_0-\delta g/2}{2}\mu_B B_m & \frac{g_0+\delta g/2}{2}\mu_B B_m & 0\\[2mm]
\frac{g_0-\delta g/2}{2}\mu_B B_m & -\frac{J}{4}+\delta g(\frac{\hbar\Omega}{2g_0}+\frac{\mu_B\delta B(t)}{2}) & \frac{J}{2} & \frac{g_0+\delta g/2}{2}\mu_B B_m\\[2mm]
\frac{g_0+\delta g/2}{2}\mu_B B_m & \frac{J}{2} &
-\frac{J}{4}-\delta g(\frac{\hbar\Omega}{2g_0}+\frac{\mu_B\delta
B(t)}{2}) &  \frac{g_0-\delta g/2}{2}\mu_B B_m\\[2mm]
0 & \frac{g_0+\delta g/2}{2}\mu_B B_m & \frac{g_0-\delta
g/2}{2}\mu_B B_m & \frac{J}{4}-g_0\mu_B\delta B(t)
\\\end {array} \right ). \eqno {(1)}$$
\end{widetext}

\begin{figure}
\includegraphics[width=3.2in]{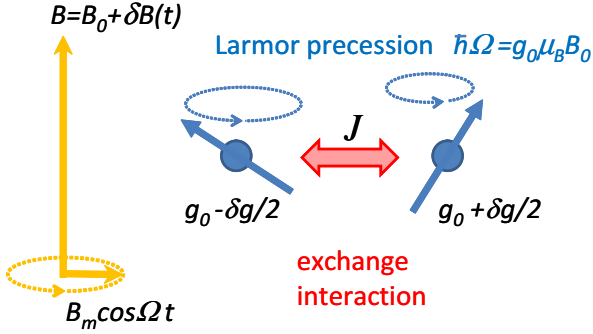}
\caption{\label{f2} The proposed model of the two-qubit system
allowing the implementation of basic one- and two-qubit operations
using a time-dependent global magnetic field $B(t)$. }
\end{figure}

\section {Realization}
In this Section we describe the strategy for finding an appropriate
control function $\delta B(t)$ for realization of desired one- or
two-qubit operations. All operations are considered in the reference
frame rotating with the frequency $\Omega$ around axis $Z$.

Let us introduce an ``ideal'' unitary operator $\mathbb{U}$ that is
defined by the desired logical operation and transforms an initial
state $|\psi_0\rangle$ of the system of two spins into a final state
$\mathbb{U} |\psi_0\rangle$.

The actual evolution of the system state vector $|\psi_t\rangle$ is
described by the equation
\[
i\hbar\frac{d|\psi_t\rangle}{dt}=\hat H(t)|\psi_t\rangle, \eqno
{(2)}
\]
where the Hamiltonian $\hat H(t)$ depends on $\delta B(t)$ according
to Eq.(1).  An unitary operator $\hat U_t$, that transforms
$|\psi_0\rangle$ into $|\psi_t\rangle$, is defined by the equation
\[
i\hbar\frac{dU_t}{dt}=\hat H(t)U_t, \eqno {(3)}
\]
and the initial condition that $U_0$ is the identity operator.

At some time $T$, the transformation $\hat U_{t=T}$ should become
equivalent to the ``ideal'' operator $\mathbb{U}$, in order  to
perform desired logical operation without an error.  In other words,
for all $|\psi\rangle$ the results of actions of $\hat U_{t=T}$ and
$\mathbb{U}$ should be physically equivalent:
\[
U_{t=T}|\psi\rangle=\mathbb{U}|\psi\rangle \, e^{i\alpha}
\]
with an arbitrary phase $\alpha$. Then the transformations
themselves are related in a similar way:
\[
U_{t=T}=\mathbb{U} \, e^{i\alpha}.
\]

The transformation $U_{t=T}$ depends on the function $\delta B(t)$
and on the duration $T$ of the logical operation. To find the best
implementation of the given quantum gate $\mathbb{U}$, we minimize
the  Frobenius norm of the deviation $\|  U_{t=T} - \mathbb{U} \,
e^{i\alpha} \| $ by variation of the function $\delta B(t)$ and
parameters $T$ and $\alpha$.

The most suitable form of the small additive $\delta B(t)$ is
\[ \delta B(t)=A\cos(\omega t+\varphi)+C \eqno {(4)}
\]
with parameters $A$, $\omega$, $\varphi$, $C$ depending on the type
of the desired logical operation. Such dependence can be easily
realized experimentally. Changing these parameters, one can perform
various one- and two-qubit quantum gates, as it will be demonstrated
below.

Fig.~3 demonstrates the example of behavior of error functional
\[f(t)=\text{min}_\alpha\| U_{t} - \mathbb{U} \, e^{i\alpha} \|^2,
\eqno {(5)}\] during the operation ``$\pi/2$-rotation around $Z$''.
It is clearly seen that the error functional has the minimum value
at $t=T$.

\begin{figure}
\includegraphics[width=3.5in]{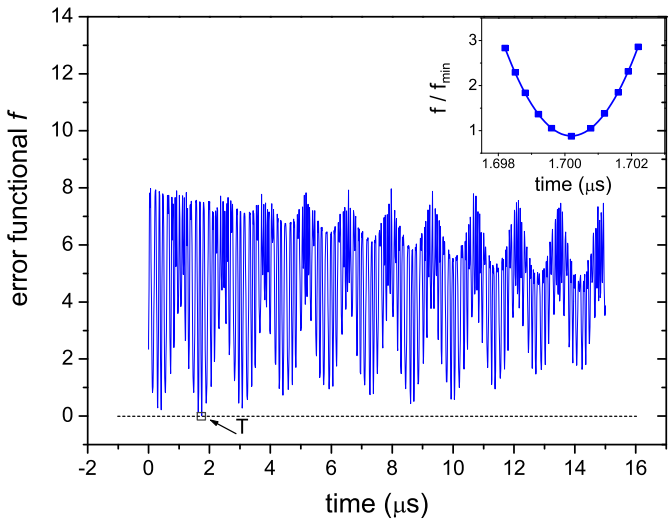}
\caption{\label{f2} Time dependence of the error functional $f(t)$,
Eq.(5), for the operation ``$\pi/2$-rotation around $Z$''. The time,
when the error functional has the minimum, is indicated as $T$.
Inset shows the tolerance range for the time $T$. One can see that
the end of logical gate can be controlled with accuracy $\pm2$~ns,
at that the increase of error $f$ does not exceed threshold
$3f_{min}$.}
\end{figure}

One can also use a more complicated modulated function $\delta B(t)$
having zero values at the beginning and at the end of the control
action:
\[ \delta B(t)=A\sin(\omega_0 t)(\cos(\omega t+\varphi)+C),\eqno {(6)}
\]
where $\omega_0=2\pi/T$ with $T$ being the time of gate
implementation. Such a form of the control function is convenient
from the standpoint of the implementation of a large sequence of
logical operations and smooth interfacing between operations. The
search for optimal parameters ($\omega_0$, $A$, $\omega$, $\varphi$,
$C$) of the function (6) can be done in two stages. The first step
is the solution of the problem with a control function described by
Eq.~(4) to determine the time $T$ at which the error of gate
implementation is minimal. At the second step one can apply the
function described by Eq.~(6) with $\omega_0=2\pi/T$ to refine the
solution.

In the following Section we will demonstrate that at reasonable
values of tunneling coupling and g-factor difference one can realize
a minimal set of one-  and two-qubit operations required for quantum
computation. As we mentioned above, the small difference of
g-factors $\delta g\sim10^{-3}$ observed recently in Ge/Si system
with quantum dots \cite{a17} is suitable for our purposes.  The
magnitude of the exchange interaction can be estimated as
$J\sim4I^2/U\sim10^{-10}$~eV, where $I$ is the tunneling integral,
$I\sim10^{-3}$~meV, and $U$ is the Coulomb interaction,
$U\sim30$~meV. The microwave magnetic field can be taken as
$B_m=1$~G. The time-dependent magnetic fields create the electric
fields, but their magnitude is rather small. The alternating
magnetic field with an amplitude $\sim1$~G causes the electric field
of the order of 300 V/cm. On the quantum dot length scale
($\sim10$~nm) the induced voltage is about 0.3 mV.  Compared with
the electron confinement potential in QD system ($\sim100$~meV),
this value is negligible, and we do not take its effect into
account.

\begin{figure}
\includegraphics[width=\linewidth]{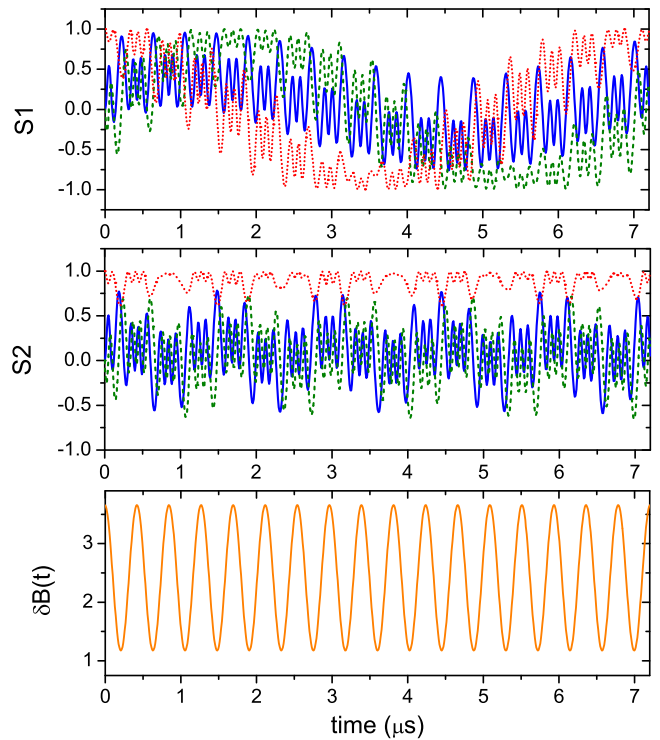}
\caption{\label{f2} Time evolution of spin components of the first
(top panel) and the second electron (center panel) during ``Storage
operation''. The blue solid curves are related to the
$S_x$-components of the electron spins $\mathbf{S}_1$ and
$\mathbf{S}_2$, the green dashed curves are related to the
$S_y$-components, the red dotted curves are related to the
$S_z$-components. The control magnetic field $\delta B(t)$ is shown
in the bottom panel. }
\end{figure}

\section {Results and discussion}
In Table~I the time $T$ and the value of gate fidelity $F$ are given
for all verified logical operations. The gate fidelity is the
probability of obtaining the correct result, averaged over all
initial states \cite{a3,a29}. The connection between the error
functional $f$ and fidelity $F$ is described in Appendix A. For all
realizations we use the function $\delta B(t)$ described by Eq.(4).
Parameters $A$, $\omega$, $\varphi$, $C$ were chosen to minimize the
deviation of the fidelity from 1, using the optimization methods
described in Ref.~\onlinecite{a12}.

\begin{table*}
\caption{\label{f2}  Results of numerical experiments. The
controlling magnetic field has the form  $\delta B(t)=A\cos(\omega
t+\varphi)+C$. The value $(1-F)$ gives the averaged probability of
the error, $F$ is the gate fidelity. The results were obtained for
the following parameters: mean g-factor value $g_0=2$, g-factor
difference $\delta g=1.1\cdot10^{-3}$, exchange interaction
$J=10^{-10}$~eV, microwave
 frequency $\Omega=9\cdot10^{9}$~Hz, microwave field amplitude  $B_m=1$~G. }
\begin{tabular}{|c|c|r|r|r|r|r|} \hline
Operation & $1-F$ & $\omega$~(MHz) & $A$~(G) & $\varphi$ & $C$~(G) & $T$($\mu$s) \\
\hline\hline Storage & $1.95\cdot10^{-5}$ & $14.80$ & $1.24$ & $0.08$& $2.417$ & $7.212$ \\
\hline

$\sqrt{SWAP}$ & $9.44\cdot10^{-4}$ & $6.79$ & 0.99 & 3.14 & $-0.009$ & 11.104 \\
\hline

$SWAP$ & $1.06\cdot10^{-3}$ & $6.79$  & $0.87$  & $0.00$ & $-0.005$& 22.209 \\
\hline

$\pi/2$-rotation
around $Z$& $4.00\cdot10^{-3}$ & $14.39$ & $0.81$ & $0.85$ & $2.319$ & $1.700$  \\
\hline

$\pi/2$-rotation around $X$ & $3.64\cdot10^{-3}$ & $17.43$
& 1.96  & 6.27 & $-2.710$ & $9.375$ \\
\hline

$\pi/2$-rotation around $Y$ & $4.53\cdot10^{-3}$ & $14.40$
& 1.24  & 1.82 & $1.323$ & $1.309$ \\
\hline

$\pi/4$-rotation around $X$ & $1.40\cdot10^{-3}$ & $9.741$ & $0.07$
& 3.05 & $0.253$ & $0.664$
\\ \hline

$\pi/8$-rotation around $X$ & $5.55\cdot10^{-3}$ & $4.816$ & $1.00$
& 5.99 & $-0.029$ & $1.424$
\\ \hline

\end{tabular}
\end{table*}

Fig.~4 demonstrates the results  for the operation ``quantum
information storage''. The top panel shows the evolution of spin
components $S_x$, $S_y$, $S_z$ for the first electron.  For the
second electron the spin evolution is shown at the central panel.
And the bottom panel demonstrates the time dependence of the
magnetic field $\delta B(t)$, that provides the conservation of the
system in the initial state. It is clearly seen that the system
under the action of the controlling magnetic field returns to its
initial state. The different behavior is observed for the first and
the second electrons, the first electron besides high-frequency
oscillations of spin components has low-frequency ones, while the
second electron has only high-frequency oscillations.  This
difference is explained by nonzero constant $C$ in the function
$\delta B(t)$, that shifts one of the electrons closer to the
resonance. In principle, it is possible to realize the identical
behavior of the spins with  $C=0$, certainly in this case the time
$T$ and other parameters of the function $\delta B(t)$ will be
changed.

In Fig.~5 the evolution of  the electron spins without
time-dependent controlling magnetic field is shown. Results
demonstrate that it is impossible to find such a moment of time, at
which both electron spins return to their initial orientations.
\begin{figure}
\includegraphics[width=3.4in]{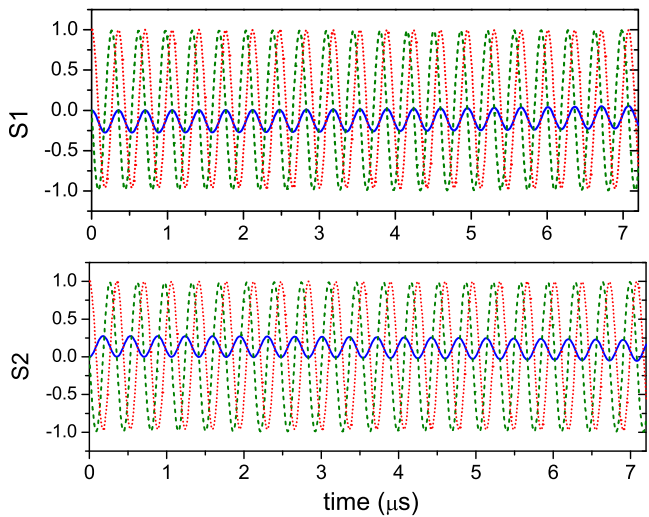}
\caption{\label{f2} Free evolution of spin components ($\delta B=0$)
of the first (top panel) and the second electron (bottom panel). The
system does not return to the initial state. }
\end{figure}

\begin{figure}
\includegraphics[width=3.2in]{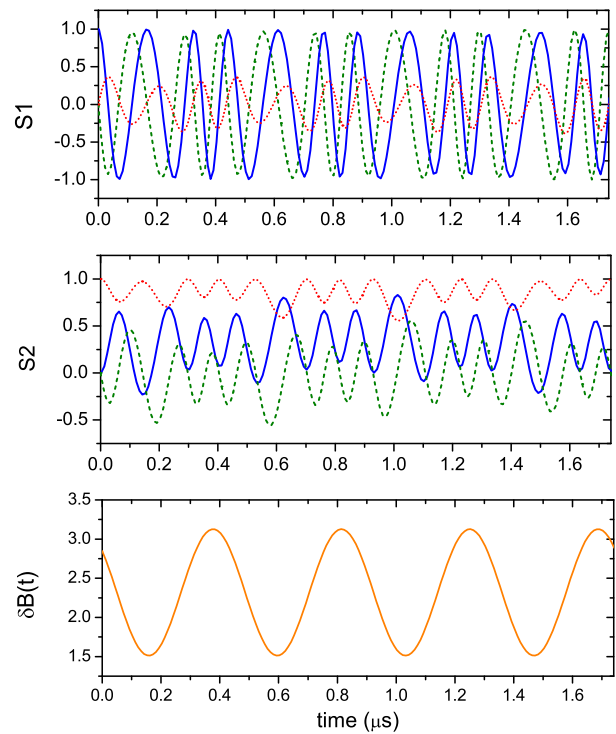}
\caption{\label{f2} Time evolution of the spin components of the
first (top panel) and the second electron (center panel) during the
operation ``$\pi/2$-rotation around $Z$''. The blue solid curves are
related to $S_x$-components of electron spins $\mathbf{S}_1$ and
$\mathbf{S}_2$, the green dashed curves are related to
$S_y$-components, the red dotted curves are related to
$S_z$-components. The control magnetic field $\delta B(t)$ is shown
in the bottom panel. }
\end{figure}

The evolution of electron spins for the operation ``$\pi/2$-rotation
around $Z$'' is presented in Fig.~6 for the following initial
conditions. The first electron spin is oriented along the axis $X$,
the second electron spin is directed along axis $Z$. As a result of
the operation, the spin of the first electron has arrived to the
orientation along the axis $Y$, while the second electron spin has
returned to its initial orientation. This demonstrates the
correctness of the proposed gate implementation.

\begin{figure}
\includegraphics[width=3.2in]{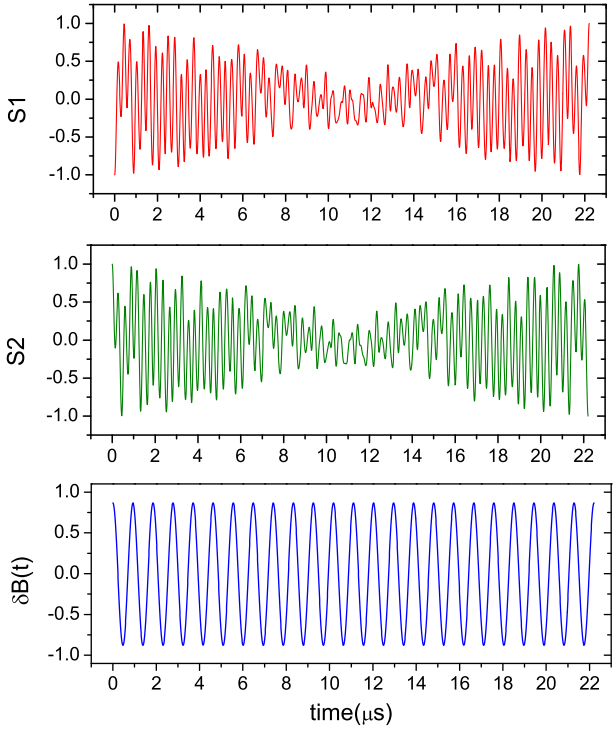}
\caption{\label{f2} Time evolution of $S_z$-components of the first
(top panel) and the second electron (center panel) during the $SWAP$
operation. The control magnetic field $\delta B(t)$ is shown in the
bottom panel.}
\end{figure}

Fig.~7 demonstrates the system evolution for the  $SWAP$ operation.
This operation is implemented if the initial state transforms to the
final state in the following manner:
\begin{gather*}
  |\!\!\uparrow\uparrow\rangle \rightarrow |\!\!\uparrow\uparrow\rangle \; e^{i\alpha}, \\
  |\!\!\uparrow\downarrow\rangle \rightarrow |\!\!\downarrow\uparrow\rangle \; e^{i\alpha}, \\
  |\!\!\downarrow\uparrow\rangle \rightarrow |\!\!\uparrow\downarrow\rangle \; e^{i\alpha}, \\
  |\!\!\downarrow\downarrow\rangle \rightarrow |\!\!\downarrow\downarrow\rangle \; e^{i\alpha},
\end{gather*}
where horizontal arrows represent the transformation from
$|\psi_0\rangle$ to $|\psi_T\rangle$ under the given Hamiltonian. In
Fig.~7 the first electron was initially in the spin-down state, and
the second one was in the spin-up state. As result of operation the
electrons have exchanged their spin directions.

Simple reduction of the operation time by a factor of 2,
$T=T_{SWAP}/2$, produces the $\sqrt{SWAP}$ gate, which has then the
maximum entangling capability. The results presented in Table I show
that the frequency $\omega$ is the same for the $\sqrt{SWAP}$ and
$SWAP$ gates, when the amplitude $A$ and the phase $\varphi$ are
slightly different.

The error probability $(1-F)$ for the spin rotation gates is found
in the range $(1.4\div 5.5)\cdot10^{-3}$, while the error
probability for the $SWAP$ operation is approximately $10^{-3}$.  We
suppose that the main source of errors in single qubit operations
(spin rotations) is a closeness of electron $g$-factors. To verify
this hypothesis we perform numerical experiments for the operation
``$\pi/2$-rotation around $X$'' with twofold increased $\delta
g=2.2\cdot10^{-3}$. As a result we obtain the five-fold reduction of
the error, $(1-F)\approx10^{-3}$.

A quality of the $SWAP$ operation depends on two parameters:
exchange interaction $J$ and detuning $\delta g/2$ of the electron
g-factors from the resonance. If the magnitude of the exchange will
be very large, the evolution of the two-electron system  becomes
less controllable. For example, the numerical experiment for the
$SWAP$ gate with $J=10^{-9}$~eV (at $\delta g=1.1\cdot10^{-3}$)
gives the error $(1-F)=3.52\cdot10^{-3}$. Regarding the parameter
$\delta g$, its increase also leads to the  error rise. In
particular, the $SWAP$ operation with $\delta g=2.2\cdot10^{-3}$
($J=10^{-10}$~eV) has the error $(1-F)=1.73\cdot10^{-3}$. In the
last case the detuning of electrons from resonance is too strong and
it becomes hard to use the spin rotations for control (or
elimination) of the exchange interaction. In Fig.~9 the behavior of
the error $(1-F)$ with change of $\delta g$ and $J$ is shown for the
$SWAP$ operation. It is clearly visible that the values of $\delta
g=1.1\cdot10^{-3}$ and $J=10^{-10}$~eV used in our numerical
experiments are close to the optimal ones. It should be noted that
we chose these values based on the parameters of real Ge/Si quantum
dots. From the dependence shown in the left panel of Fig.~9 it seems
that the value $\delta g=5\cdot10^{-4}$ is more appropriate  than
$\delta g=1.1\cdot10^{-3}$. However, the decrease of $\delta g$
complicates the addressing individual qubits, that provokes the rise
of the error $(1-F)$ during single-qubit operations. Thus, the
search for the optimal values of $\delta g$ and $J$ should be
carried out with respect  to the whole set of one- and two-qubit
operations.

\begin{figure}
\includegraphics[width=3.2in]{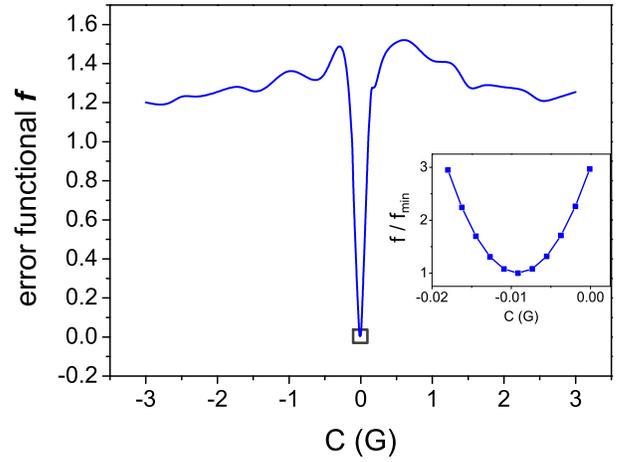}
\caption{\label{f2} Dependence of the error functional $f$ on the
deviation of the parameter $C$ (a constant shift in the controlling
field $\delta B(t)=A\cos(\omega t+\varphi)+C$) for the $\sqrt{SWAP}$
gate. Inset shows that at $\Delta C\leq0.01$~G the error $f$ does
not exceed the threshold $3f_{min}$.  }
\end{figure}

\begin{figure}
\includegraphics[width=\linewidth]{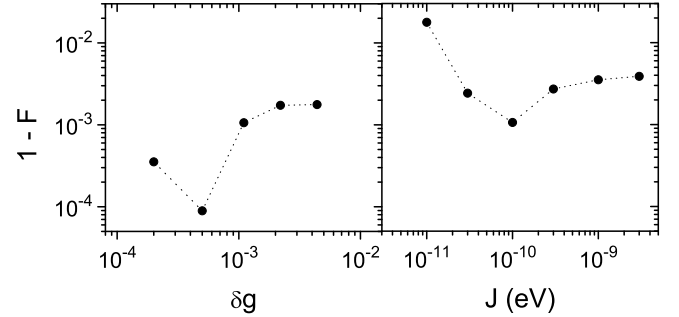}
\caption{\label{f2} Dependence of the error $(1-F)$  on the g-factor
difference $\delta g$ (left panel) and on the exchange integral $J$
(right panel) for the $SWAP$ gate. Here $F$ is the gate fidelity.}
\end{figure}

It is worth noting that each optimization problem considered in this
paper can have a few possible solutions with approximately equal
errors. In principle one can use any of these solutions. Each
solution has its own unique moment of time, when the system reaches
the desired state. Since the behavior of the error functional $f$
represents very strong oscillations, the success of our approach
depends  on the accuracy of time determination. The tolerable error
$\Delta t$, at that the averaged probability of error $(1-F)$ stays
the same order of magnitude, is about $\pm 2$~ns (Fig.~3). Therefore
the gate duration should be controlled with accuracy $\simeq1$~ns.
Such a requirement can be easily satisfied with modern experimental
equipment.

Additionally we have studied the reliability of our approach with
respect to deviation of parameters of the controlling field $\delta
B(t)$ ($A$, $\omega$, $\phi$, $C$) from their optimal values. The
study was performed for the $\sqrt{SWAP}$ operation. In a real
experiment all these parameters can be controlled with some
accuracy, and we have found for each parameter the tolerance range
at which the fidelity $F$ remains of the same order as the optimal
value. We define the tolerance range from the condition $f\leq
3f_{min}$, where $f_{min}$ is the error functional value given in
Table I. In Fig.~8 the tolerance range for the parameter $C$ is
shown. From the inset to this figure one can see that the tolerance
range $\Delta C=\pm0.01$~G. However, the fidelity is more sensitive
to the amplitude $A$. The deviation $\Delta A =\pm0.001$~G causes
the increase of the error $f$ up to $3f_{min}$. The frequency
$\omega$ must be controlled with accuracy $\Delta\omega=\pm5$~kHz.
The tolerant change of the phase $\varphi$ is
$\Delta\varphi=\pm0.12$. It should be noted that within the
indicated tolerance ranges the time of the gate implementation does
not change.

Finally we would like to discuss the experimental feasibility of the
magnetic field control with required accuracy. The  magnetic field
in our proposal consists of two parts $B + \delta B(t)$. The first
large part is of the order of $10^3$~G, the second small part is
about 1~G. Both fields should be stabilized with accuracy
$10^{-3}$~G. The most challenging problem is stabilization of the
large magnetic field, because the relative accuracy should be of the
order of $10^{-6}$. Modern superconducting magnets are able to meet
this requirement. For example, the specification of uniformity of
the magnetic fields in the general NMR apparatus is 0.01 ppm in a
sample space and 0.01 ppm/hour  in time-based stability. The
additive part $\delta B(t)$ can be produced by a small extra coil,
and should be controlled with a reasonable accuracy $\sim0.1\%$.

\section{Conclusions }
We have proposed a new method of the implementation of one- and
two-qubit gates in a system of two electrons with constant exchange
coupling. The only controlled parameter is the small time-dependent
addition to the magnetic field. Small g-factor difference between
two electrons provides not only the selective access to the
individual qubits, but also the effective control of the exchange
interaction between them. We have verified the possibility of basic
one- and two-qubit operations in the frame of proposed model and
found the parameters of controlling magnetic field allowing to
perform these operations with an acceptable accuracy.

\begin{acknowledgments}
This work was supported by RFBR (Grant 13-02-12105), SB RAS
integration project No. 83 and DITCS RAS project No. 2.5.
\end{acknowledgments}

\appendix
\section{relation between error functional and fidelity}

In this Appendix, we will show how to find the gate fidelity $F$ from
the functional $f$ defined by Eq.~(5).

Frobenius norm of a square matrix $A$ is defined as
\[
 \| A \| = \sqrt{\text{Tr}\left(AA^+\right)} ,
\]
therefore
\begin{multline*}
 \left\| U_t - \mathbb{U} \, e^{i\alpha} \right\|^2 =
 \text{Tr} \left[ \left( U_t-\mathbb{U}\,e^{i\alpha} \right)
 \left( U^+_t-\mathbb{U}^+\,e^{-i\alpha} \right) \right]
 \\
 = \text{Tr}\left(U_tU_t^+\right)
 + \text{Tr}\left(\mathbb{U}\mathbb{U}^+\right)
 - 2\,\text{Re}\left[ e^{i\alpha}\, \text{Tr}\left(U_t^+\mathbb{U}\right) \right] .
\end{multline*}
In the latter expression, $U_tU_t^+$ and $\mathbb{U}\mathbb{U}^+$
are identity matrices $4\times4$, so their traces are equal to 4.
Denoting the remaining trace as $Z$,
\[
  Z = \text{Tr}\left(U_t^+\mathbb{U}\right) ,
\]
one obtains the following relation:
\[
 \left\| U_t - \mathbb{U} \, e^{i\alpha} \right\|^2 =
 8 - 2\,\text{Re}\left( e^{i\alpha}\,Z \right) .
\]
Hence, according to Eq.~(5),
\begin{equation}
  f(t) = \text{min}_\alpha  \left\| U_t - \mathbb{U} \, e^{i\alpha} \right\|^2 =
 8 - 2\,|Z| .
\end{equation}

Since the matrix $U_t^+\mathbb{U}$ is unitary, its four eigenvalues can be expressed as
$e^{i\varphi_1},\ldots,e^{i\varphi_4}$, where $\varphi_1,\ldots,\varphi_4$ are real numbers.
So the trace $Z$ of this matrix is equal to
\[
  Z = \sum_{k=1}^4 e^{i\varphi_k} ,
\]
therefore
\begin{multline}
  |Z|^2 = \sum_k \sum_l e^{i\varphi_k} e^{-i\varphi_l} \\
  = 4 + 2\sum_{k<l} \cos(\varphi_k-\varphi_l)
  \equiv 4+2M ,
\end{multline}
where
\[
  M = \sum_{k<l} \cos(\varphi_k-\varphi_l) .
\]
Using Eqs.~(A1) and~(A2), one can express the quantity $f$ via $M$:
\begin{equation}
  f = 8-2\sqrt{4+2M} .
\end{equation}

Now let us express the gate fidelity $F$ via $M$. The gate fidelity is the probability,
averaged over all initial state vectors $|\psi\rangle$, that the system
will be found in the desired state $\mathbb{U}|\psi\rangle$ after passing through the gate:
\[
  F = \overline{\left|\langle\psi|U_t^+\mathbb{U}|\psi\rangle\right|^2} ,
\]
where the overline denotes averaging over initial states. Let us consider this expression
in a basis that diagonalizes the matrix $U_t^+\mathbb{U}$:
\[
  U_t^+\mathbb{U} = \left ( \begin {array}{cccc}
  e^{i\varphi_1}& 0& 0&0\\[2mm]
  0 & e^{i\varphi_2}&0&0 \\[2mm]
  0 & 0 & e^{i\varphi_3}&0\\[2mm]
  0 & 0 & 0&e^{i\varphi_4}\\[2mm]
  \end {array} \right ),
  \quad
  |\psi\rangle = \left ( \begin {array}{c}
  c_1\\[2mm]
  c_2 \\[2mm]
  c_3\\[2mm]
  c_4\\[2mm]
  \end {array} \right ),
\]
where $c_1,\ldots,c_4$ are complex amplitudes that define the state $|\psi\rangle$.
In this basis,
\[
  \langle\psi|U_t^+\mathbb{U}|\psi\rangle = \sum_{k=1}^4 |c_k|^2 e^{i\varphi_k} ,
\]
therefore
\begin{multline*}
 F = \sum_k \sum_l \overline{|c_k|^2|c_l|^2} e^{i\varphi_k} e^{-i\varphi_l} \\
 = \sum_k \overline{|c_k|^4} + 2 \sum_{k<l} \overline{|c_k|^2|c_l|^2} \cos(\varphi_k-\varphi_l) .
\end{multline*}
It is possible to show that
\[
  \overline{|c_k|^2|c_l|^2} = \begin{cases}
  1/10 \quad \text{if } k=l, \\
  1/20 \quad \text{if } k\not=l,
  \end{cases}
\]
therefore
\begin{equation}
  F = \frac{4+M}{10} \,.
\end{equation}
Comparing Eqs.~(A3) and~(A4), one can easily see that
\begin{equation}
  F = 1 - \frac{f}{5} \left(1-\frac{f}{16}\right) .
\end{equation}
If $f$ is small enough, then $(1-F)$ is approximately proportional to $f$:
\[
  1-F \approx \frac{f}{5} .
\]
Hence, the minimization of the error functional $f$ is equivalent to
the minimization of the deviation of the fidelity $F$ from~1.


\begin{thebibliography}{24}

\bibitem{a1}
T. D. Ladd, F. Jelezko, R. Laflamme, Y. Nakamura, C. Monroe, and J.
L. O'Brien, {\em Nature} {\bf 464},  45 (2010).

\bibitem{a2}
D. P. DiVincenzo, in {\em Mesoscopic Electron Transport}, eds. Sohn,
Kowenhoven, Schoen (Kluwer 1997), p. 657, cond-mat/9612126; ÓThe
Physical Implementation of Quantum Computation,Ô Fort. der Physik
{\bf 48}, 771 (2000), quant-ph/0002077

\bibitem{a3}
D. Loss and D. P. DiVincenzo, Phys. Rev. A {\bf 57}, 120 (1998).

\bibitem{a4}
L. P. Kouwenhoven, C. M. Marcus, P. L. McEuen, S. Tarucha, R. M.
Westervelt  and N. S. Wingreen.   {\em Mesoscopic Electron Transport
Series E},  ed. L. L. Sohn, L. P. Kouwenhoven and G. Sch\"{o}n
(Dordrecht: Kluwer), {\bf 345}, 105 (1997).

\bibitem{a5}
W. G. Van der Wiel, S. De Franceschi, J. M. Elzerman, T. Fujisawa,
S. Tarucha  and L. P. Kouwenhoven,   Rev. Mod. Phys. {\bf 75} 1
(2003).

\bibitem{a6}
H.~A.~Engel  and D.~Loss,   Phys. Rev. Lett. {\bf 86}, 4648 (2001).

\bibitem{a7}
R. Vrijen, E. Yablonovitch, K. Wang, H. W. Jiang, A. Balandin, V.
Roychowdhury, T. Mor, D. DiVincenzo, Phys. Rev. A {\bf 62}, 012306
(2000).

\bibitem{a8}
R. Brunner, Y.-S. Shin, T. Obata, M. Pioro-Ladriere, T. Kubo, K.
Yoshida, T. Taniyama, Y. Tokura, and S. Tarucha, Phys. Rev. Lett.
{\bf 107}, 146801 (2011).

\bibitem{a9}
Y. Hu, Z.-W. Zhou, G.-C. Guo, New Journal of Physics {\bf 9}, 27
(2007).

\bibitem{a10}
X. Zhou, Z.-W. Zhou, G.-C. Guo, and M. J. Feldman, Phys. Rev. Lett.
{\bf 89}, 197903 (2002).

\bibitem{a11}
M. A. Nielsen, I. L. Chuang {\it Quantum Computation and Quantum
Information} (Cambridge University Press, New York, 2010).

\bibitem{a12}
Y. Ozhigov, L. Fedichkin, JETP Letters {\bf 77}, 328 (2003).

\bibitem{a13}
C.A. Floudas and P.M. Pardalos (Eds.), {\it Encyclopedia of
Optimization} (Second edition, Springer, 2009).

\bibitem{a14}
A. Abragam,  {\it The principles of nuclear magnetism} (Oxford:
Clarendon Press, 1961).

\bibitem{a15}
E. L. Ivchenko and A. A. Kiselev, Sov. Phys. Semicond. {\bf 26}, 827
(1992).

\bibitem{a16}
H. Kosaka, A. A. Kiselev, F. A. Baron, K. W. Kim, and E.
Yablonovitch, Electron. Lett. {\bf 37}, 464 (2001).

\bibitem{a17}
A. F. Zinovieva,  A. I. Nikiforov, V.A. Timofeev, A. V. Nenashev, A.
V. Dvurechenskii,  L. V. Kulik, Phys. Rev. B {\bf 88}, 235308
(2013).

\bibitem{a18}
J. Levy, Phys. Rev. Lett. {\bf 89}, 147902 (2002).

\bibitem{a19}
J. M. Taylor, H.-A. Engel, , W. Du\"{e}, A. Yacoby, C. M. Marcus, P.
Zoller, M. D. Lukin Nature Physics {\bf 1}, 177 (2005).

\bibitem{a20}
N. Yokoshi, H. Imamura, H. Kosaka, Phys. Rev. B {\bf 81}, 161305
(2010).

\bibitem{a21}
M. Grydlik, G. Langer, T. Fromherz, F. Sch\"{a}ffler, M. Brehm,
Nanotechnology {\bf 24},  105601 (2013).

\bibitem{a22}
V. A. Zinovyev, A. V. Dvurechenskii, P. A. Kuchinskaya, and V. A.
Armbrister, Phys. Rev. Lett. {\bf 111}, 265501 (2013).

\bibitem{a23}
G.~Feher, Phys. Rev. {\bf 114}, 1219 (1959); G. Feher  and  E. Gere,
Phys. Rev. {\bf 114}, 1245 (1959).

\bibitem{a24}
M. Chiba  and A. Hirai, J. Phys. Soc. Jpn.  {\bf 33}, 730 (1972).

\bibitem{a25}
A. F. Zinovieva, A. V. Dvurechenskii, N. P. Stepina, A. I.
Nikiforov, A. S. Lyubin, L. V. Kulik,  Phys. Rev. B.  {\bf 81},
113303 (2010).

\bibitem{a26}
A. F. Zinovieva, N. P. Stepina, A. I. Nikiforov, A. V. Nenashev, A.
V. Dvurechenskii, L. V. Kulik, M. C. Carmo, and N. A. Sobolev, Phys.
Rev. B {\bf 89}, 045305 (2014).

\bibitem{a27}
A. F. Zinovieva, A. V. Nenashev, A. V. Dvurechenskii, Phys. Rev. B
{\bf 71}, 033310 (2005).

\bibitem{a28}
E. Abe, A. M. Tyryshkin, S. Tojo, J. J. L. Morton, W. M. Witzel, A.
Fujimoto, J. W. Ager, E. E. Haller, J. Isoya, S. A. Lyon, M. L. W.
Thewalt, K. M. Itoh, Phys. Rev. B {\bf 82}, 121201(R) (2010).

\bibitem{a29}
J. F. Poyatos, J.-I. Cirac, and P. Zoller, Phys. Rev. Lett. {\bf
78}, 390 (1997).




\end{thebibliography}
\end{document}